\documentclass[prd,nofootinbib,showpacs]{revtex4}

\usepackage{graphicx}
\usepackage{dcolumn}
\usepackage{bm}
\usepackage{mathrsfs,amsmath}
\usepackage{hyperref}
\usepackage{color}
\usepackage{subfigure}

\newcommand{\beq}{\begin{equation}}
\newcommand{\eeq}{\end{equation}}

\usepackage{listings}

\newcommand{\code}[1]{\lstinline|#1|} 

\begin{document}

\title{CosmicFish Implementation Notes V1.0}

\author{Marco Raveri$^{1,2,3}$, Matteo Martinelli$^{4,5}$, Gong-Bo Zhao$^{6,7}$ and Yuting Wang$^{6,7}$}
\affiliation{
\smallskip
$^{1}$ SISSA - International School for Advanced Studies, Via Bonomea 265, 34136, Trieste, Italy \\
\smallskip
$^{2}$ INFN, Sezione di Trieste, Via Valerio 2, I-34127 Trieste, Italy \\
\smallskip
$^{3}$ INAF-Osservatorio Astronomico di Trieste, Via G.B. Tiepolo 11, I-34131 Trieste, Italy \\
\smallskip
$^{4}$ Institute Lorentz, Leiden University, PO Box 9506, Leiden 2300 RA, The Netherlands \\
\smallskip
$^{5}$ Institut f\"ur Theoretische Physik, Ruprecht-Karls-Universit\"at Heidelberg, Philosophenweg 16, 69120 Heidelberg, Germany.  \\
\smallskip
$^{6}$ National Astronomy Observatories, Chinese Academy of Science, Beijing, 100012, P.R.China \\
\smallskip
$^{7}$ Institute of Cosmology \& Gravitation, University of Portsmouth, Dennis Sciama Building, Portsmouth, PO1 3FX, UK
}

\begin{abstract}
CosmicFish is a publicly available library to perform Fisher matrix forecast for several cosmological observations.
With the present implementation notes we provide a guide to the physical and technical details of the library.
We reproduce here the details and all the relevant equations, as they appear in the code.
We submit these notes to the arXiv to grant full and permanent access to this material which provides a useful guidance to forecasting and the use of CosmicFish code.
We will update this set of notes when relevant modifications to the CosmicFish code will be released.
The present version is based on CosmicFish Jun16.
\end{abstract}

\date{\today}


\maketitle

\tableofcontents

\newpage 

\section{Introduction and overview}
These notes contain the CosmicFish implementation details of all the physical quantities relevant to the code.
The core of the CosmicFish code consists of two libraries. The first one is a Fortran library that takes care of producing the Fisher matrices. This is interfaced with CAMB sources~\cite{Lewis:1999bs, Challinor:2011bk}, EFTCAMB sources~\cite{Hu:2013twa, Raveri:2014cka} and MGCAMB sources~\cite{Zhao:2008bn, Hojjati:2011ix} and is automatically compiled against these three Einstein-Boltzmann codes for maximum model coverage. \\
The second part consists of a Python library optimized to perform operations on Fisher matrices, once they are produced. This also contains a full set of plotting utilities.
Both these two libraries have applications built with them. For the Fortran part the only application already present is one that computes Fisher matrices. Other applications are in the process of being developed.
The Python part contains several plotting applications that produce 1D, 2D and triangular marginalized plots, an application producing tables with marginalized bounds on the parameters and an application that performs a complete analysis of a set of Fisher matrices.
Both libraries are based on state of the art, optimized, core algorithms including precise derivatives calculators, spectral protection of Fisher matrices against degenerate, unconstrained, parameters, to name a few. 

The CosmicFish code is publicly available at~\url{http://cosmicfish.github.io}. \\
The code comes with two unit testing suites that are responsible for checking all the building blocks of the two libraries. These can be easily used to check whether the code is correctly deployed and functioning or if modification introduced to the code break the correct execution of previous features. \\
The two libraries constituting the CosmicFish code are supplied with a thorough automatic documentation that explain the interfaces and the purposes of all the functions in both codes. This documentation is accessible at~\url{http://cosmicfish.github.io/documentation/CosmicFish/index.html} for the Fortran part and at~\url{http://cosmicfish.github.io/documentation/CosmicFishPyLib/index.html} for the Python part. \\
The codes comes with two packages that we used to validate the code and to perform the analysis presented in~\cite{CosmicFishIG}. These packages are built to fully exploit the power of the library and show an example of a work pipeline with the CosmicFish library. The structure of these packages is extremely flexible and they can be easily used to build packages with other applications of the code. \\
The CosmicFish code is also publicly released in a developers version containing the latest update and features that are not yet implemented in the stable version.
Thanks to its flexible and documented structure the developers version of the CosmicFish code can be easily modified to implement new features that can be useful to the cosmological community. We encourage everyone that is interested in developing its own applications or modifications to the code to use this version of the code, exploit it for scientific purposes and then to upload modifications to it to join our efforts in pursuing cosmological forecasting as a community achievement. \\
We believe that, in its first release, the CosmicFish code stands as a flexible and powerful framework to produce cosmological forecasts. Its fully modular structure, modern design and utilities make it a perfect tool to produce transparent and accessible scientific results and represent a first step toward community wide forecasting efforts.
\section{Contributing to the CosmicFish code development}
Cosmological forecasting is intrinsically a complicated matter as it requires deep knowledge of many branches of cosmology. 
In particular performing useful forecast requires modeling extended physical theories, understanding systematic effects related to cosmological observables and experimental design.
All this can hardly be mastered by a single individual. For this reason we believe that CosmicFish, as a forecasting code, has to be open to contributions from other people that will bring their expertise to the code. \\
The developers version of the CosmicFish code is available at~\url{https://github.com/CosmicFish/CosmicFish} and we welcome anyone to submit contributions. \\
In an effort toward allowing scientific recognition of contributors to the code we shall use the following scheme. \\
All contributions, regardless of size, are recognized and acknowledged in one of the CosmicFish files, distributed with the main code and in the acknowledgement section of these notes. \\
Contributors to the code are further divided into three levels:
\begin{itemize}
\item {\it Contributors}: anyone contributing to the code;
\item {\it Developers}: anyone contributing to the CosmicFish code in the long-term and that provided substantial additions to the code;
\item {\it Principal Developers}: anyone that contributed to the code to the same extent of other Principal Developers and with a global overview of the code;
\end{itemize}
To help developers in justifying their effort in contributing to the CosmicFish code, as an academic undertaking, we shall prepare a scientific paper exploiting the new features of the code at any mayor release of its stable version. {\it Principal Developers} shall be invited to sign the paper, {\it Developers} shall be invited to contribute to the realization of the paper. This paper will accompany the release of the new version of the code and will be the paper for which the code license requires a citation by the users. In addition {\it Principal Developers} will be invited to sign the CosmicFish implementation notes.

\section{Fisher forecasts}
Given a data set $D$ and a model $\mathcal{M}$ that is described by a vector of parameters $\theta$  we call the likelihood of the data $\mathcal{L}(\theta) \equiv P(D|\theta|\mathcal{M})$ the probability of the data given a set of parameters and a model. \\
The Fisher information matrix is then defined by:
\begin{align} \label{Eq:FisherMatrixDefinition}
F_{ij} \equiv \bigg\langle \left( \frac{\partial \ln \mathcal{L} }{\partial \theta_i}\right) \left( \frac{\partial \ln \mathcal{L} }{\partial \theta_j}\right) \bigg\rangle_{D} = \int \left( \frac{\partial \ln \mathcal{L} }{\partial \theta_i}\right) \left( \frac{\partial \ln \mathcal{L} }{\partial \theta_j}\right) \mathcal{L}\, dD
\end{align}
Where brackets denote average over data realizations.
The Fisher matrix can also be written as:
\begin{align} \label{Eq:FisherMatrixSecondDefinition}
F_{ij} \equiv -\bigg\langle \left( \frac{\partial^2 \ln \mathcal{L} }{\partial \theta_i \partial \theta_j}\right) \bigg\rangle_{D} = - \int \left( \frac{\partial^2 \ln \mathcal{L} }{\partial \theta_i \partial \theta_j}\right) \mathcal{L}\, dD
\end{align}
since the difference between the two is zero once averaged over the data $D$. \\
The relevance of this quantity comes from to the Cram\'er-Rao lower bound~\cite{KayStat}. This states that if:
\begin{align}
\bigg\langle \frac{\partial \ln  \mathcal{L} }{\partial \theta_i}\bigg\rangle_{D} = \int \left( \frac{\partial \ln  \mathcal{L} }{\partial \theta_i} \right) \mathcal{L}\, dD = 0 \hspace{0.5cm} \forall \theta_i \,\,,
\end{align}
the covariance $C$ of any unbiased estimator $\hat{\theta}$ of the parameters satisfies:
\begin{align} \label{Eq:CramerRao}
C(\hat{\theta}) - F^{-1} \geq 0
\end{align}
where $\geq 0$ means that this quantity is positive semidefinite. \\

\section{Statistical analysis of Fisher matrices}
In this section we discuss some of the statistical analyses that can be performed with Fisher matrices at hand. These are all implemented in the CosmicFish code and we refer to the code documentation,~\url{http://cosmicfish.github.io/documentation/CosmicFish/index.html} and~\url{http://cosmicfish.github.io/documentation/CosmicFishPyLib/index.html}, for the implementation details.
\subsection{Marginalized bounds}
Once Fisher matrices are produced we can use them to forecast marginal bounds on parameters.
Given a Fisher matrix $F$, the Cram\'er-Rao lower bound~(\ref{Eq:CramerRao}) implies that:
\begin{align}
{\rm var} ( \hat{\theta_i} ) \geq \left( F^{-1} \right)_{ii}
\end{align}
If we assume that the posterior of the considered parameter is Gaussian we can forecast the bound at different confidence levels:
\begin{align}
{\rm C.L. \, bound} = \sqrt{2} \, {\rm Erfinv}\left( {\rm C.L.} \right) \sqrt{ \left( F^{-1} \right)_{ii} } \hspace{0.5cm}{\rm where}\hspace{0.5cm} 0\leq {\rm C.L.} \leq 1 \,,
\end{align}
where ${\rm Erfinv}$ denotes the inverse Error function and ${\rm C.L.}$ is the desired confidence level.

\subsection{Marginal ellipses plotting}
Once we compute a Fisher matrix it might be useful to plot the marginalized joint posterior probability of two parameters $(\theta_i, \theta_j)$, assuming that it is Gaussian, starting from the Fisher matrix. This helps shedding light on the possible degeneracy between two parameters and how this degeneracy changes or breaks when considering different experiments.
Let us denote:
\begin{align}
\sigma_{i}^2 =& \left( F^{-1} \right)_{ii} \nonumber \\ 
\sigma_{j}^2 =& \left( F^{-1} \right)_{jj} \nonumber \\ 
\sigma_{ij}^2 =& \left( F^{-1} \right)_{ij}
\end{align}
we can immediately compute:
\begin{align}
a =& \sqrt{ \frac{1}{2}( \sigma_{i}^2 + \sigma_{j}^2 ) + \sqrt{ \frac{1}{4} ( \sigma_{i}^2 - \sigma_{j}^2 )^2 + \sigma_{ij}^4 }  } \nonumber \\
b =& \sqrt{ \frac{1}{2}( \sigma_{i}^2 + \sigma_{j}^2 ) - \sqrt{ \frac{1}{4} ( \sigma_{i}^2 - \sigma_{j}^2 )^2 + \sigma_{ij}^4 }  }  \nonumber \\
\phi_0 =& \frac{1}{2} {\rm atan} \left( \frac{ 2 \sigma_{ij}^2 }{ \sigma_{i}^2 - \sigma_{j}^2} \right) 
\end{align}
and we can write the parametric form of the Fisher ellipse in the $\theta_i, \theta_j$ plane as:
\begin{align}
\theta_i =& \, \alpha \left( a \cos(\phi)\cos(\phi_0) - b \sin(\phi)\sin(\phi_0) \right) + \theta_i^0 \nonumber \\
\theta_j =& \, \alpha \left( a \cos(\phi)\sin(\phi_0) + b \sin(\phi)\cos(\phi_0) \right) + \theta_j^0 \hspace{0.5cm} \forall \phi \in [0,2\pi]
\end{align}
where $(\theta_i^0, \theta_j^0)$ is the value of the fiducial parameters and $\alpha$ is a coefficient encoding the confidence level of the ellipse and is given by $\alpha=\sqrt{2} \, {\rm Erfinv}\left( {\rm C.L.} \right)$ where ${\rm Erfinv}$ denotes the inverse Error function and ${\rm C.L.}$ is the desired confidence level $0\leq {\rm C.L.} \leq 1$.

\subsection{Information Gain}
We can use Fisher matrices to forecasts the information gain between different experiments~\cite{Seehars:2014ora} as we shall review in this section.
The statistical tool that we shall use at this goal is the Kullback-Leibler divergence, also called relative entropy or information gain.
Consider two probability density functions (PDF), $P_1$ and $P_2$ of a $d$ dimensional random variable $\theta$. The Kullback-Leibler (KL) divergence is defined by:
\begin{align} \label{Eq:KLDivergence}
D\left( P_2 || P_ 1 \right) \equiv \int P_2(\theta) \log_{2}\left(\frac{P_2(\theta)}{P_1(\theta)} \right)  \, d\theta = \frac{1}{\ln(2)}\int P_2(\theta) \ln\left(\frac{P_2(\theta)}{P_1(\theta)} \right)  \, d\theta \hspace{1cm} \mbox{[ bits ]}
\end{align}
and represents the information difference in going from $P_1$ to $P_2$ in bits.
The KL divergence has some noticeable properties:
\begin{itemize}
\item {\it Positive definite}: $D\left( P_2 || P_ 1 \right)\geq 0$ and $D\left( P_2 || P_ 1 \right)=0$ iff $P_1=P_2$;
\item {\it Not symmetric}: $D\left( P_2 || P_ 1 \right)\neq D\left( P_1 || P_ 2 \right)$;
\item {\it Invariant under re-parametrizations}: given $Y(X)$, a non-singular re-parametrization, $D\left( P_2(Y) || P_ 1(Y) \right)=D\left( P_2(X) || P_ 1(X) \right)$.
\end{itemize}
If we assume that $P_1$ and $P_2$ are multivariate Gaussian distributions with mean $\theta_1$ and $\theta_2$ and covariance $\Sigma_1$ and $\Sigma_2$ the KL divergence becomes:
\begin{align} \label{Eq:KLDivergenceGaussian}
D\left( P_2 || P_ 1 \right) = \frac{1}{2\ln 2}\left( \theta_1 - \theta_2 \right)^{T} \Sigma_1^{-1} \left( \theta_1 - \theta_2 \right) + \frac{1}{2\ln 2}\left[ -\ln\frac{\det \Sigma_2}{\det \Sigma_1} -d +{\rm Tr} \left( \Sigma_2\Sigma_1^{-1} \right)\right]
\end{align}
We can consider the posterior of some experiments as the PDF in the KL divergence to quantify the information difference between the interpretation of two different experiments within a model.
With reference to Eq. (\ref{Eq:KLDivergence}) we shall consider $P_1= P(\theta | D_1, \mathcal{M} ) = P(\theta | \mathcal{M}) P( D_1 | \theta, \mathcal{M}) \equiv P(\theta) \mathcal{L}( D_1, \theta)$ and $P_2= P(\theta | D_2, \mathcal{M} ) = P(\theta | \mathcal{M}) P( D_2 | \theta, \mathcal{M}) \equiv P(\theta) \mathcal{L}( D_2, \theta)$ where $D_1$ and $D_2$ are two data sets and $\mathcal{M}$ is a specific model. \\
If we apply the KL divergence to the posterior of two different experiments sometimes it proves useful to compute the expectation and variance of the KL divergence over $D_2$ realizations:
\begin{align}
\langle D\left( P_2 || P_ 1 \right) \rangle_{D_2} \equiv& \int D\left( P_2 || P_ 1 \right)  \mathcal{L}( D_2 ) \, d D_2 \label{Eq:KLDivergenceExpectation}\\
\sigma^2(D) \equiv& \int \left[ D\left( P_2 || P_ 1 \right) -\langle D\left( P_2 || P_ 1 \right) \rangle \right]^2  \mathcal{L}( D_2 ) \, d D_2 \label{Eq:KLDivergenceVariance}
\end{align}
We shall now consider the case where the two posterior are Gaussian in the parameters and the data and have the same mean values. We shall denote the Fisher matrix of the two experiments as $F_1$ and $F_2$ and the prior Fisher matrix as $F_p$. With these assumptions and notation it can be easily shown~\cite{Seehars:2014ora} that:
\begin{align}
D\left( P_2 || P_ 1 \right)  =& \frac{1}{2\ln 2} \left[ -\ln \frac{\det \left( F_1+F_p\right) }{\det \left( F_2+F_p \right)} -d +{\rm Tr}\left[ \left( F_2+F_p\right)^{-1} \left( F_1+F_p \right) \right]\right] \label{Eq:KLDivergenceFisher} \\
\langle D\left( P_2 || P_ 1 \right) \rangle  =& \frac{1}{2\ln 2} \left[ -\ln \frac{\det \left( F_1+F_p\right) }{\det \left( F_2+F_p \right)} -d +{\rm Tr}\left[ \left( F_2+F_p\right)^{-1} \left( F_1+F_p \right) \right]\right] \nonumber \\
	&+\frac{1}{2\ln 2} {\rm tr} \bigg[ F_2\left( F_p + F_2 \right)^{-1} \left( F_p + F_1 \right) \left( F_p + F_2 \right)^{-1} \left( I +F_2 \left( F_p + F_1 \right)^{-1} \right) \bigg]  \label{Eq:KLDivergenceExpectedFisher}\\
\sigma^2(D) =& \frac{1}{2\ln 2} {\rm tr} \bigg\{ \left[ F_2\left( F_p + F_2 \right)^{-1} \left( F_p + F_1 \right) \left( F_p + F_2 \right)^{-1} \left( I +F_2 \left( F_p + F_1 \right)^{-1} \right) \right]^2\bigg\}   \label{Eq:KLDivergenceVarianceFisher}
\end{align}
Equations~(\ref{Eq:KLDivergenceFisher}, \ref{Eq:KLDivergenceExpectedFisher}, \ref{Eq:KLDivergenceVarianceFisher}) are the results that are implemented in the CosmicFish code to perform information gain forecasts.\\
Some properties of~(\ref{Eq:KLDivergenceFisher}, \ref{Eq:KLDivergenceExpectedFisher}, \ref{Eq:KLDivergenceVarianceFisher}) are worth noticing:
\begin{itemize}
\item {\it KL divergence with uninformative priors}: if the prior is less informative than the Fisher matrix, i.e. $F_1, F_2 >> F_p$ then:
\begin{align}
D\left( P_2 || P_ 1 \right)  =& \frac{1}{2\ln 2} \left[ -\ln \frac{\det F_1}{\det  F_2 } -d +{\rm Tr}\left[ F_2^{-1} F_1 \right]\right] \nonumber  \\
\langle D\left( P_2 || P_ 1 \right) \rangle  =&  \frac{1}{2\ln 2} \left[ -\ln \frac{\det F_1}{\det  F_2 } +2{\rm Tr}\left[ F_2^{-1} F_1 \right]\right] \nonumber \\
\sigma^2(D) =& \frac{1}{2\ln 2} {\rm tr} \bigg[ F_1F_2^{-1}F_1F_2^{-1} + 2F_1F_2^{-1}\bigg]  +\frac{d}{2\ln 2}
\end{align}
\item {\it KL divergence with strong prior}: in the limit where the prior is much more informative than the data, i.e. $F_p >> F_1, F_2$ we have:
\begin{align}
\lim_{F_p >> F_1, F_2} D\left( P_2 || P_ 1 \right) \rightarrow 0 \hspace{0.2cm};\hspace{0.2cm} \langle D\left( P_2 || P_ 1 \right) \rangle \rightarrow 0 \hspace{0.2cm};\hspace{0.2cm} \sigma^2(D) \rightarrow 0
\end{align}
that is, if the prior are too strong the information that we can gain is reduced to zero.
\end{itemize}

\section{Fisher matrix for angular power spectra}
In this section we shall describe how we obtain the Fisher matrix for angular power spectra~\cite{Tegmark:1996bz}. \\
Consider $N$ fields measured on the sky. These will constitute a vector $\vec{T}$ of angle $(\theta)$ dependent measurements that we can decompose into spherical harmonics:
\begin{align}
\vec{T}(\theta) = \sum_{l,m} \vec{a}_{lm} Y_{lm}(\theta) \hspace{0.5cm}\mbox{with}\hspace{0.5cm} {\rm dim}(\vec{a}_{lm}) =N
\end{align}
We shall assume that the $\vec{a}_{lm}$ are Gaussian distributed with zero mean and covariance $C_{l}^{XY} = \langle | a_{lm}^X a_{lm}^{Y+} |\rangle$ where $X$ and $Y$ denote two of the observables in $\vec{T}$. 
This covariance constitutes our theoretical prediction and we shall compare it to the data estimate of the covariance: 
\begin{align}
\hat{C}_{l} = \frac{1}{2l+1} \sum_{m=-l}^l \vec{a}_{lm} \vec{a}_{lm}^+
\end{align}
At a fixed $m$ and $l$ the probability of $\vec{a}_{lm}$ can be written as:
\begin{align}
P(\vec{a}_{lm} | C_l ) = (2\pi)^{-N/2} |C_l^{XY}|^{-1/2} \exp \left[ -\frac{1}{2} \vec{a}_{lm}^T \left( C_l^{XY} \right)^{-1} \vec{a}_{lm} \right]
\end{align}
where we used the notation $\det(\cdot) = |\cdot|$.
The joint probability distribution of the $2l+1$ independent $\vec{a}_{lm}$ is then given by:
\begin{align}
P(\vec{a}_{lm} | C_l ) =& \prod_{m=-l}^l \left[  (2\pi)^{-N/2}  |C_l^{XY}|^{-1/2} \exp \left[ -\frac{1}{2} \vec{a}_{lm}^T \left( C_l^{XY} \right)^{-1} \vec{a}_{lm} \right] \right] \nonumber \\
=& \left( 2 \pi \right)^{-N(2l+1)/2} |C_l^{XY}|^{-(2l+1)/2} \exp \left[ -\frac{2l+1}{2} \sum_{XY} \hat{C}_l^{XY}\left( C_l^{XY} \right)^{-1}  \right]
\end{align}
An experiment measures the $C_l$'s from $l_{\rm min}$ to $l_{\rm max}$ so that the joint probability distribution of the full measurements is given by:
\begin{align}
\ln \mathcal{L} = \ln P( \hat{C}_l | C_l ) =& -\frac{1}{2} N\log\left( 2\pi\right)\left( 1+l_{\rm max} - l_{\rm min} \right)\left( 1+l_{\rm max} + l_{\rm min} \right) \nonumber \\
&-\frac{1}{2}\sum_{l=l_{\rm min}}^{l_{\rm max}} \left( 2l+1\right) \bigg[ \log |C_l^{XY}| +{\rm Tr}\left[   \hat{C}_l^{XY}\left( C_l^{XY} \right)^{-1} \right] \bigg]
\end{align}
With this at hand we can immediately compute the Fisher matrix for angular power spectra as:
\begin{align}
F_{ij} = \sum_{l=l_{\rm min}}^{l_{\rm max}} \frac{2l+1}{2} \sum_{\alpha \beta \gamma \delta = 1}^{N} \frac{\partial C_l^{\alpha \beta } }{\partial \theta_i} \left(\hat{C}_l^{\beta \gamma}\right)^{-1} \frac{\partial C_l^{\gamma \delta} }{\partial \theta_i} \left(\hat{C}_l^{\delta \alpha}\right)^{-1} 
\end{align}
Here we model the observed $\hat{C}_l^{XY}$ as:
\begin{align}
\hat{C}_l^{XY} = \frac{\left( C_l^{XY} + N_{l}^{XY} \right)}{ \sqrt[4]{f_{\rm sky}^X f_{\rm sky}^Y}}
\end{align}
The angular power spectra that are already implemented in the CosmicFish code are:
\begin{itemize}
\item CMB temperature, E and B mode polarization, CMB lensing as discussed in Section~\ref{SubSec:CMBSpectra}.
\item Galaxy weak lensing, as discussed in Section~\ref{SubSec:LSS};
\item Galaxy number counts fluctuations, as discussed in Section~\ref{SubSec:LSS};
\end{itemize}
To compute the Fisher matrix for angular power spectra the parameter flag \code{cosmicfish_want_cls} has to be turned to true.
By default the cross correlation between all the angular power spectra is considered. To remove cross-correlation from the Fisher matrix calculation the user should act on the flag \code{Fisher_want_XC}.

\subsection{CMB angular power spectra} \label{SubSec:CMBSpectra}
We implement four different CMB angular power spectra accounting for their cross correlation.
In particular we have CMB Temperature spectrum, CMB E-mode polarization spectrum, CMB B-mode polarization spectrum and CMB lensing angular spectrum.
Since we consider only statistical noise these are not correlated and the noise in the cross-correlation is zero. 
In the case of the Temperature spectrum we consider different frequency channels $\nu$ so that the total statistical noise is given by:
\begin{align} \label{Eq:TemperatureNoise}
N_{\ell}^{TT} =& \left( \sum_{\nu} \frac{1}{\rm var(\nu)} \exp \bigg ( -\ell (\ell+1) \sigma^2(\nu) \bigg)  \right)^{-1} \nonumber \\
{\rm var}(\nu) =& \left( S_T(\nu) \theta(\nu) \frac{\pi \, T_{\rm CMB} }{180 \times 60} \right)^2 \nonumber \\
\sigma^2(\nu) =& \left( \frac{\pi \, \theta(\nu) }{180 \times 60 \sqrt{8 \log 2} } \right)^2
\end{align}
where $\theta(\nu)$ is the beam FWHM in arcmin and $S_T(\nu) \equiv \Delta T / T$ is temperature sensitivity.
In the case of E-B mode polarization the noise is given by the same Eq.~\ref{Eq:TemperatureNoise} with $S_T(\nu)$ replaced by $S_P(\nu) \equiv \Delta P / T$ as polarization sensitivity. \\
In the case of CMB lensing noise is given by~\cite{Hu:2001kj} and computed with the code developed in~\cite{Perotto:2006rj}.
Different CMB spectra can be included in the Fisher matrix calculation by acting on the flags: \code{Fisher_want_CMB_T}, \code{Fisher_want_CMB_E}, \code{Fisher_want_CMB_B} and \code{Fisher_want_CMB_lensing}.
The relevant parameters for CMB experiments are:
\begin{itemize}
\item \code{CMB_n_channels}: number of frequency channels of the CMB experiment;
\item \code{CMB_TT_fsky}, \code{CMB_EE_fsky} and \code{CMB_BB_fsky} to select $f_{\rm sky}$ for different observables;
\item \code{l_max_TT}, \code{l_max_EE} and \code{l_max_BB} to select the maximum multipole for a given observable;
\item \code{CMB_temp_sens(i)} to select temperature sensitivity ($\Delta T / T$) in the \code{i}-th frequency channel;
\item \code{CMB_pol_sens(i)} to select polarization sensitivity ($\Delta P / T$) in the \code{i}-th frequency channel;
\item \code{CMB_fwhm(i)}  to select the beam FWHM in arcmin for the \code{i}-th frequency channel;
\end{itemize}

\subsection{LSS angular power spectra} \label{SubSec:LSS}

We implement in the CosmicFish code the possibility to include Large Scale Structure (LSS) angular power spectra in the Fisher matrix forecast.
In particular, based on the output of CAMB sources, we implement galaxy Weak Lensing (WL) and tracers number counts fluctuations, hereafter GC. \\
As with CMB spectra, by default the code includes in the forecast all the cross correlation between different LSS observables and also the cross correlation between CMB observables and LSS observables. \\
In this section we discuss the details of the implementation of LSS forecast.

\subsubsection{Window functions}
For LSS observables we need to specify the corresponding window functions. We implement several choices that we shall discuss in this section.

\begin{itemize}
\item {\it Gaussian window function}: the Gaussian window function is specified as:
\begin{align}\label{Eq:GaussianWindow}
W(z) = \frac{1}{\sqrt{2\pi} \sigma} \exp \left( -\frac{1}{2}\frac{\left( z-z_0\right)^2}{\sigma^2}\right)
\end{align}
where $z_0$ indicates the window mean redshift and $\sigma$ denotes the window spread. This window can be selected with the parameter \code{window_type=1}. The parameters of this window can be specified to the code by: \code{redshift(i)}$=z_0$ in the \code{i}-th window; \code{redshift_sigma(i)}$=\sigma$ in the \code{i}-th window. 
\item {\it Binned window function}: the binned window function is specified by a global window function that encodes the information about all the objects found by a survey. We implement:
\begin{align} \label{Eq:GlobalBinWindow}
\bar{N}_{\rm survey}(z) = \frac{\beta}{z_0^{(1+\alpha)} \Gamma\left(\frac{1+\alpha}{\beta} \right)}z^\alpha \exp\left[ -\left(\frac{z}{z_0}\right)^\beta \right]
\end{align}
where $\alpha$, $\beta$ and $z_0$ are the parameters specifying the window function. In particular $\alpha$ and $\beta$ are two parameters defining the shape of the window function, $\Gamma$ is the Euler gamma function and $z_0$ is the window median redshift.
This global window function has several properties:
\begin{align} \label{Eq:GlobalBinWindowProperties}
\int_{z_1}^{z_2} \bar{N}_{\rm survey}(z) \,dz &= \frac{1}{\Gamma\left(\frac{1+\alpha}{\beta} \right)} \left[ \Gamma\left(\frac{1+\alpha}{\beta}, \left(\frac{z_1}{z_0}\right)^\beta \right) - \Gamma\left(\frac{1+\alpha}{\beta}, \left(\frac{z_2}{z_0}\right)^\beta \right) \right] \nonumber \\
\int_{0}^{\infty} \bar{N}_{\rm survey}(z) \,dz &= 1
\end{align}
Photometric redshift determinations have always some errors. To account for them the window function is usually smoothed.
If we consider a redshift bin going from $z_1$ to $z_2$ the Gaussian smoothed binned window function becomes:
\begin{align} \label{Eq:BinWindow}
W(z) &= \int_{z_1}^{z_2} \bar{N}_{\rm survey}(z) \exp \left[ -\frac{\left( \tilde{z} - z \right)^2}{2\sigma_z^2(z)}\right] \, d\tilde{z} = \bar{N}_{\rm survey}(z) \int_{z_1}^{z_2} \exp \left[ -\frac{\left( \tilde{z} - z \right)^2}{2\sigma_z^2(z)}\right] \, d\tilde{z} \nonumber \\
&= \frac{\bar{N}_{\rm survey}(z)}{2}\left[ {\rm Erf}\left(\frac{z-z_1}{\sqrt{2}\sigma_z(z)} \right) - {\rm Erf}\left(\frac{z-z_2}{\sqrt{2}\sigma_z(z)}  \right) \right]  \nonumber \\
&= \frac{\bar{N}_{\rm survey}(z)}{2}\left[ {\rm Erf}\left(\frac{z-z_1}{\sqrt{2}\sigma_z(z)} \right) - {\rm Erf}\left(\frac{z-(z_1+\Delta z)}{\sqrt{2}\sigma_z(z)}  \right) \right] 
\end{align}
Where $\sigma_z^2(z)$ is the model for the photometric redshift error and ${\rm Erf}$ is the error function.
We implement a simple model for $\sigma_z$ that is specified by $\sigma_z(z) = \sigma_0(1+z)$.
The parameters for this window function can be specified to the code by:
\begin{itemize}
\item \code{window_type=2} to select this kind of window;
\item \code{window_alpha}$=\alpha$;
\item \code{window_beta}$=\beta$;
\item \code{redshift_zero}$=z_0$;
\item \code{photoz_error}$=\sigma_0$;
\item \code{redshift(i)}$=z_1$ in the \code{i}-th window;
\item \code{redshift_sigma(i)}$=\Delta z$ in the \code{i}-th window;
\end{itemize}
\item {\it Flat window function}: this window function is specified by:
\begin{align} \label{Eq:FlatWindow}
W(z) = \left\{
\begin{array}{ll}
      1 & z_1 \leq z \leq z_1+\Delta z \\
      0 & \mbox{otherwise} \\
\end{array} 
\right.
\end{align}
This window function can be selected by setting \code{window_type=3} and the parameters: \code{redshift(i)}$=z_1$ in the \code{i}-th window; \code{redshift_sigma(i)}$=\Delta z$ in the \code{i}-th window. 
\item {\it Smoothed flat window function}: this window function is equivalent to the flat window function but accounting for photo-z error Gaussian smoothing. It is specified by:
\begin{align} \label{Eq:FlatSmoothWindow}
W(z) &= \frac{\bar{N}_{\rm survey}(z)}{2}\left[ {\rm Erf}\left(\frac{z-z_1}{\sqrt{2}\sigma_z(z)} \right) - {\rm Erf}\left(\frac{z-(z_1+\Delta z)}{\sqrt{2}\sigma_z(z)}  \right) \right] 
\end{align}
where:
\begin{align}
\bar{N}_{\rm survey}(z) = \left\{
\begin{array}{ll}
      1 & z_1 \leq z \leq z_1+\Delta z \\
      0 & \mbox{otherwise} \\
\end{array} 
\right.
\end{align}
This window function can be selected by setting \code{window_type=4} and the parameters: \code{redshift(i)}$=z_1$ in the \code{i}-th window; \code{redshift_sigma(i)}$=\Delta z$ in the \code{i}-th window. We implement $\sigma_z(z) = \sigma_0(1+z)$ where $\sigma_0=$\code{photoz_error}.
\end{itemize}

Notice that all these window functions are automatically normalized to unity by the CAMB code so we do not need to specify the normalization.

These window functions can be used for both GC and WL. In its current version, the code does not allow to use two different functional forms of window functions at the same time. This is likely to be upgraded in the future.

All the LSS windows have also the possibility of having a different $f_{\rm sky}$ and a different $l_{\rm max}$ by setting: \code{LSS_fsky(i)} for the sky fraction and \code{LSS_lmax(i)} for the maximum multipole, in the \code{i}-th window.

\subsubsection{LSS Weak Lensing Noise}
In this section we shall discuss the experimental noise that is added to the predicted $C_{l}$. In the case of WL we consider a noise given by:
\begin{align} \label{Eq:WLNoise}
N_{\ell}^{ {\rm Wl}_i {\rm Wl}_i } = \frac{\gamma_{\rm rms}^2}{N_i}
\end{align}
where is the rms shear stemming from the intrinsic ellipticity of the galaxies and $N_i$ is the number of galaxies in the window.
In code notation: \code{LSS_intrinsic_ellipticity(i)}$=\gamma_{\rm rms}$ and \code{LSS_num_galaxies(i)}$=N_i$ in the \code{i}-th window.
\subsubsection{LSS Number Counts Noise}
In this section we shall discuss the experimental noise that we consider for number counts fluctuations. This is given by:
\begin{align} \label{Eq:GCNoise}
N_{\ell}^{ {\rm GC}_i {\rm GC}_i } = \frac{1}{N_i}
\end{align}
where $N_i$ is the number of galaxies in the window.
In code notation: \code{LSS_num_galaxies(i)}$=N_i$ in the \code{i}-th window.

For GC we need to specify bias. In the present version of the code this is treated as a constant, scale independent, possibly different in all redshift windows.
The value of bias in the \code{i}-th window is specified by: \code{redshift_bias(i)}. 
It is possible to add these values of the bias to the Fisher matrix calculation by setting \code{param[bias]} to true. More complicated bias models will be added in the near future. 

\section{Fisher matrix for Supernovae observations}
In this section we discuss how we obtain Fisher matrix forecast for Supernovae observations.
We observe the magnitude of a supernova at a certain redshift and we model it as:
\begin{align}\label{Eq:SNMagnitude}
m(z_i) = 5 \log_{10} \left( D_L(z_i) \right) -\alpha X_1 +\beta C +M_0 -25
\end{align}
where $X_1$ is the stretch of the supernova, $C$ its color and $M_B$ the intrinsic luminosity.
If we measure the luminosity distance, redshift, color and stretch of a set of $N$ supernovae and we assume that all these measurements are Gaussian distributed with mean $m(z_i)$ and covariance $\Sigma$ we can immediately write the likelihood as:
\begin{align} \label{Eq:SNLikelihood}
\mathcal{L} =  \left( 2\pi \right)^{-N/2} \det\left( \Sigma \right)^{-1/2} \exp\left( -\frac{1}{2}\left( \vec{m}_{\rm obs} - \vec{m} \right)^{\rm T}\Sigma^{-1} \left( \vec{m}_{\rm obs} - \vec{m} \right) \right)
\end{align}
We model the covariance as diagonal and:
\begin{align} \label{Eq:SNCovariance}
\Sigma = \sigma^2_m +\sigma_{\rm lens}(z)^2 +\sigma_z(z)^2 +\beta^2\sigma_C^2(z)  +\alpha^2\sigma_{X_1}^2(z) +2\alpha \sigma_{mX_1}(z) -2\beta\sigma_{Cm}(z) -2\alpha\beta \sigma_{X_1C}(z)
\end{align}
where: $\sigma_m$ is the error in determining the SN magnitude; $\sigma_{\rm lens}$ is the error due to the lensing of the SN and we model its redshift dependence as $\sigma_{\rm lens}(z) = \sigma_{\rm lens}\,z$; $\sigma_z$ is the error in the determination of the SN redshift whose redshift dependence is modeled by $\sigma_z(z) = \frac{5 \sigma_z}{z \ln(10)}$; $\sigma_C$ is the error in the determination of SN color and we model its redshift dependence as 
$\sigma_C(z) = \sigma_{C0} + \sigma_{C2}z^2$; $\sigma_{X_1}$ is the error in the determination of the SN stretch and we model its redshift dependence as 
$\sigma_{X_1}(z) = \sigma_{X_10} + \sigma_{X_12}z^2$; $\sigma_{mX_1}$ is the error in the joint determination of stretch and magnitude of a supernova and we model its redshift dependence as $\sigma_{mX_1}(z) = \sigma_{mX_1 0} + \sigma_{mX_1 2}z^2$; $\sigma_{Cm}$ is the error in the joint determination of the color and magnitude of a supernova and we model its redshift dependence as $\sigma_{Cm}(z) = \sigma_{Cm 0} + \sigma_{Cm 2}z^2$; 
$\sigma_{X_1C}$ is the error in the joint determination of the supernova stretch and color and we model its redshift dependence as $\sigma_{X_1C}(z) = \sigma_{X_1C 0} + \sigma_{X_1C 2}z^2$. \\
With this at hand we can compute the Fisher matrix that becomes:
\begin{align}
F_{ab} = \bigg\langle \frac{\partial m_i}{\partial p_a} \Sigma^{-1}_{ij} \frac{\partial m_j}{\partial p_b} +\frac{1}{2}{\rm Tr} \left( {\bf \Sigma}^{-1} \frac{\partial {\bf \Sigma}}{\partial p_a} {\bf \Sigma}^{-1} \frac{\partial {\bf \Sigma}}{\partial p_b} \right) \bigg\rangle_{D}
\end{align}
because the covariance matrix depends on the parameters $\alpha$ and $\beta$. Notice that this quantity needs to be averaged over the data. In particular the average over the observed magnitude has already been taken into account while the average over redshift, color and stretch has to be performed.	This turns out to be challenging to do analytically. We thus allow the option to perform a Monte-Carlo average of this quantity considering color and stretch to be Gaussian distributed with zero mean and user defined spread. \\
In the present version of the code the term depending on the covariance derivative is not implemented. It will be implemented in a future code release.

\subsection{SN survey parameters}
In this section we review the parameters that can be used to tell the code the specifications of a SN survey.

\begin{itemize}
\item \code{SN_Fisher_MC_samples}: number of Monte Carlo Fisher matrix samples;
\item \code{alpha_SN}: fiducial value of the stretch coefficient. This can be added as a parameter of the Fisher matrix by setting \code{param[alpha_SN]} to true;
\item \code{beta_SN}: fiducial value of the color coefficient. This can be added as a parameter of the Fisher matrix by setting \code{param[beta_SN]} to true;
\item \code{M0_SN}: fiducial value of the SN intrinsic luminosity. This can be added as a parameter of the Fisher matrix by setting \code{param[M0_SN]} to true;
\item \code{color_dispersion}: dispersion in the colors. This fixes the variance in color of the generated mock SN catalog;
\item \code{stretch_dispersion}: dispersion in the stretch. This fixes the variance in color of the generated mock SN catalog;
\item \code{magnitude_sigma}$=\sigma_m$;
\item \code{c_sigmaz}$=\sigma_z$;
\item \code{sigma_lens_0}$=\sigma_{\rm lens}$;
\item \code{dcolor_offset}$=\sigma_{C0}$;
\item \code{dcolor_zcorr}$=\sigma_{C2}$;
\item \code{dshape_offset}$=\sigma_{X_10}$;
\item \code{dshape_zcorr}$=\sigma_{X_12}$;
\item \code{cov_ms_offset}$=\sigma_{mX_1 0}$;
\item \code{cov_ms_zcorr}$=\sigma_{mX_1 2}$;
\item \code{cov_mc_offset}$=\sigma_{Cm 0}$;
\item \code{cov_mc_zcorr}$=\sigma_{Cm 2}$;
\item \code{cov_sc_offset}$=\sigma_{X_1C 0}$;
\item \code{cov_sc_zcorr}$=\sigma_{X_1C 2}$;
\item\code{number_SN_windows}: number of redshift bin in the SN redshift distribution;
\item \code{SN_redshift_start(i)}: lower redshift if the \code{i}-th  SN redshift distribution bin;
\item \code{SN_redshift_end(i)}: higher redshift if the \code{i}-th  SN redshift distribution bin;
\item \code{SN_number(i)}: number of SN in the \code{i}-th  SN redshift distribution bin;
\end{itemize}

\section{Fisher matrix for Redshift Drift measurements}
In this section we describe how we obtain the Fisher matrix for Redshift Drift (RD) measurements.\\
The theoretical shift in the spectroscopic velocity of a source after a time interval $\Delta t$ can be modelled as
\begin{align}
\Delta v(z) = cH_0\Delta t\left[1-\frac{E(z)}{1+z}\right]
\end{align}
where $E(z)\equiv H(z)/H_0$ and $H_0$ is expressed in $s^{-1}$.\\
The error on this quantity is strongly 
dependent on the measurement strategy adopted and on the observed sources; at the moment CosmicFish follows the 
modelling of the uncertainties adopted by the European Extremely Large Telescope (E-ELT), expected to produce
measurements of this effect observing QSO absorption systems. Given this setup, the uncertainties on the 
measurements can be modelled as:
\begin{align}
\sigma_{\Delta v} = 1.35\frac{2370}{S/N}\sqrt{\frac{30}{N_{QSO}}}\left(\frac{5}{1+z}\right)^x
\end{align}
where $N_{QSO}$ is the number of observed system in the redshift bin centered in $z$, $S/N$ is the signal-to-noise ratio of the observations and $x=1.7$ for $z\leq4$ and $x=0.9$ for $z>4$.\\

Definining the likelihood for a set of $N$ measurements at redshifts $z_k$ as 
\begin{align}
 \mathcal{L} \propto \exp{\left[-\sum_{k=1}^N{\frac{(\Delta v_{\rm obs}(z_k) - \Delta v(z_k))^2}{2\sigma^2_{\Delta v}}}\right]}
\end{align}

Given this shape of the likelihood function, the Fisher matrix can be written as
\begin{align}
 F_{ij} = -\frac{\partial^2\ln{\mathcal{L}(\vec{\theta})}}{\partial\theta_i\partial\theta_j}=\sum_{k=1}^N{\frac{1}{\sigma^2_k}\frac{\partial \Delta v_k}{\partial\theta_i}\frac{\partial \Delta v_k}{\partial\theta_j}}
\end{align}

The parameters that can be used to build the redshift drift data set are

\begin{itemize}
\item \code{RD_exptype}: selects between different error modelization (only $1$ available in the present version);
\item \code{number_RD_redshifts}: number of redshift bins;
\item \code{delta_time}: time interval $\Delta t$ (in years) for which to compute the shift;
\item \code{RD_sig_to_noise}: signal to noise ratio $S/N$;
\item \code{RD_source_number(i)}: number of observed sources in the i-th redshift bin;
\item \code{RD_redshift(i)}: central redshift value of the i-th redshift bin.
\end{itemize}

\section{Fisher matrix for derived parameters}
In this section we discuss how we obtain the Jacobian matrix that is needed to propagate the bounds from a set of parameters to another.
We can easily derive the transformation of the Fisher matrix under a change in the parameters. If we denote a set of parameters by $\theta_i$ and another set of parameters by $\tilde{\theta}_i$ and taking equation~(\ref{Eq:FisherMatrixDefinition}) we immediately have:
\begin{align}
\tilde{F}_{ij} = \bigg\langle \left( \frac{\partial \ln \mathcal{L} }{\partial \tilde{\theta}_i}\right) \left( \frac{\partial \ln \mathcal{L} }{\partial \tilde{\theta}_j}\right) \bigg\rangle_{D} = \frac{\partial \theta_k}{\partial \tilde{\theta}_i}\,  \bigg\langle \left( \frac{\partial \ln \mathcal{L} }{\partial \theta_k}\right) \left( \frac{\partial \ln \mathcal{L} }{\partial \theta_m}\right)  \bigg\rangle_{D}\, \frac{\partial \theta_m}{\partial \tilde{\theta}_j} = \frac{\partial \theta_k}{\partial \tilde{\theta}_i}\, F_{km} \, \frac{\partial \theta_m}{\partial \tilde{\theta}_j}
\end{align}
similarly we have:
\begin{align} \label{Eq:FisherDerivedInverse}
F_{km} =  \frac{\partial \tilde{\theta}_i}{\partial \theta_k}\, \tilde{F}_{ij} \, \frac{\partial \tilde{\theta}_j}{\partial \theta_m} 
\end{align}
The CosmicFish code then computes numerically $J \equiv \frac{\partial \tilde{\theta}_j}{\partial \theta_m}$ and the Fisher matrix with the derived parameters can be computed as:
\begin{align}
\tilde{F} = \left( J^T F^{-1} J \right)^{-1}
\end{align}
Notice that the derived Fisher matrix satisfies the Cram\'er-Rao lower bound for the derived parameters~\cite{KayStat}.
The derived parameters Jacobian matrix is computed by the CosmicFish code if \code{cosmicfish_want_derived} is set to true.
\subsection{Derived parameters}
The CosmicFish code allows to get derived bounds on several derived parameters that we list here as a reference:
\begin{itemize}
\item The present day baryon relative density $\Omega_b$. Active if \code{param[omegab]} is true;
\item The present day CDM relative density $\Omega_c$. Active if \code{param[omegac]} is true;
\item The present day neutrino relative density $\Omega_\nu$. Active if \code{param[omegan]} is true;
\item The present day DE relative density $\Omega_\Lambda$. Active if \code{param[omegav]} is true;
\item The present day curvature relative density $\Omega_k$. Active if \code{param[omegak]} is true;
\item The present day total matter relative density $\Omega_m$. Active if \code{param[omegam]} is true;
\item $\theta_{\rm CMB}$ that measures the sound horizon at last scattering. Active if \code{param[theta]} is true;
\item Total mass of massive neutrinos $m_{\nu}$. Active if \code{param[mnu]} is true;
\item The reionization redshift $z_{\rm re}$. Active if \code{param[zre]} is true;
\item Effective number of relativistic species $n_{\rm eff}$. Active if \code{param[neff]} is true;
\end{itemize}
In addition there are a couple of tomographic derived parameters, that can be obtained at several different redshifts. These are:
\begin{itemize}
\item $\sigma_{8}(z)$ the amplitude of scalar perturbations on the scale of $8\, h^{-1}{\rm Mpc}$. Active if \code{param[sigma8]} is true;
\item $\log \mathcal{H}(z)$ the logarithm of the Hubble parameter at user specified redshifts. Active if \code{param[loghubble]} is true;
\item $\log D_{A}(z)$ the logarithm of the angular diameter distance at user specified redshifts. Active if \code{param[logDA]} is true;
\end{itemize}
Tomographic derived parameters are active if \code{FD_num_redshift} is set greater than zero. If this is the case \code{FD_redshift(i)} specifies the \code{i}-th redshift of the tomographic derived parameter.
These constitute a basic set of derived parameters. It should be easy for the user to implement new ones.

\section{Examples}
\begin{figure}[h!]
\begin{center}
\subfigure[ {\it Planck} BB forecasted constraints ]{ \includegraphics[width=0.9\textwidth]{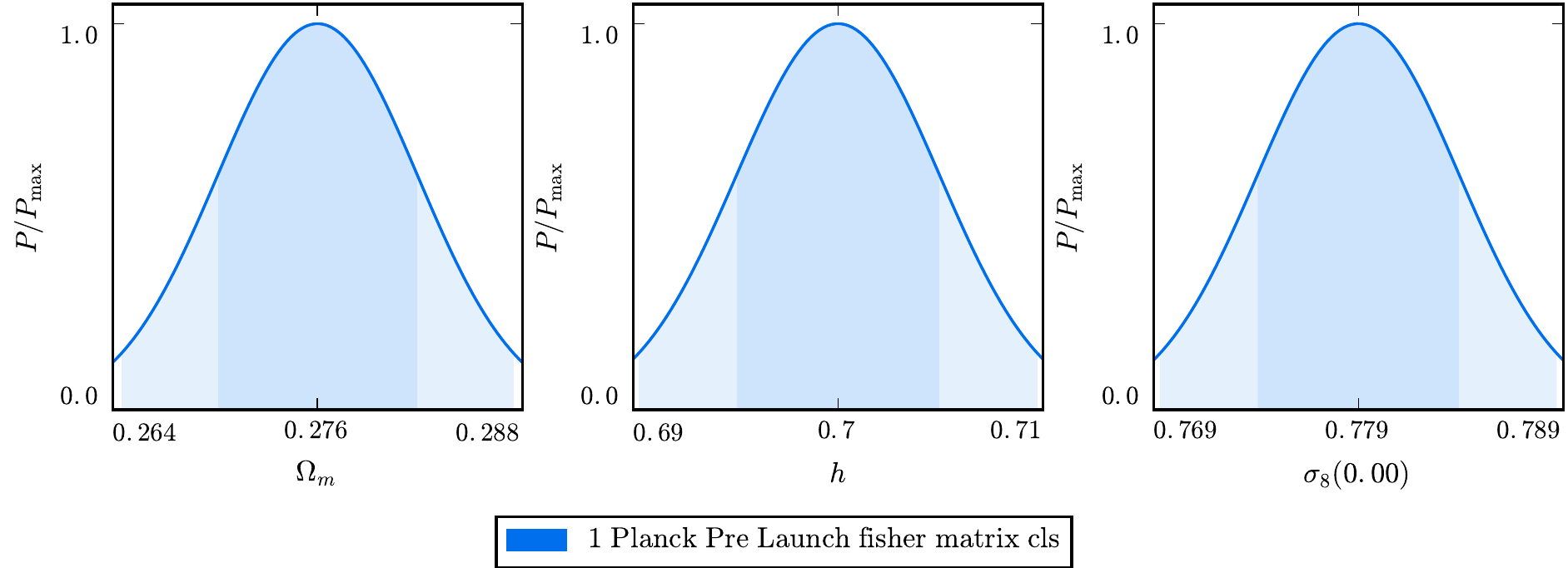} \label{SubFig:Planck1D}}
\hspace{5mm}
\subfigure[ {\it Planck} BB and DES+{\it Planck} BB forecasted constraints ]{ \includegraphics[width=0.9\textwidth]{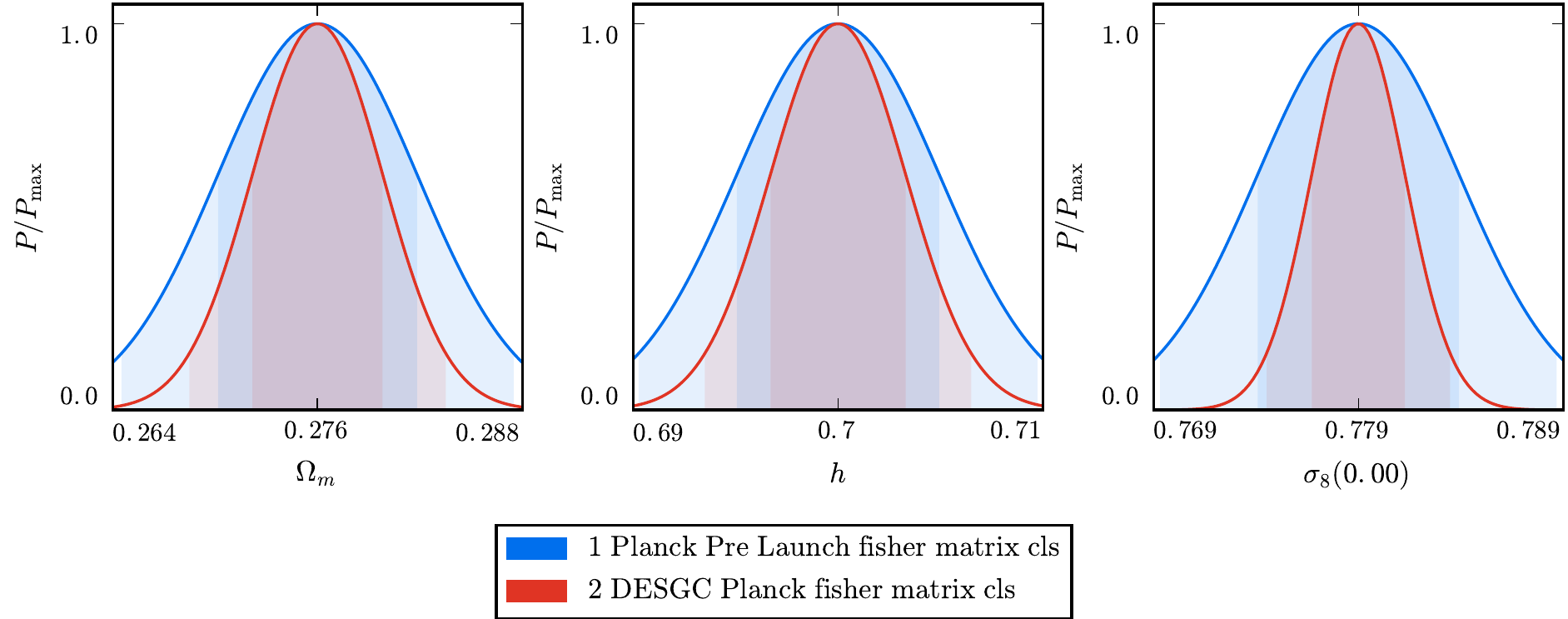} \label{SubFig:DES1D} }
\caption{The marginalized 1D forecasted constraints on $\Omega_m$, $h$ and $\sigma_8$. Different colors correspond to different combinations of experiments, as shown in legend. The darker and lighter color shades indicate the forecasted $68\%$ C.L. and $95\%$ C.L. bounds.}\label{fig:1Dfigure}
\end{center}
\end{figure}

\begin{figure}[h!]
\begin{center}
\subfigure[ {\it Planck} BB forecasted constraints ]{ \includegraphics[width=0.9\textwidth]{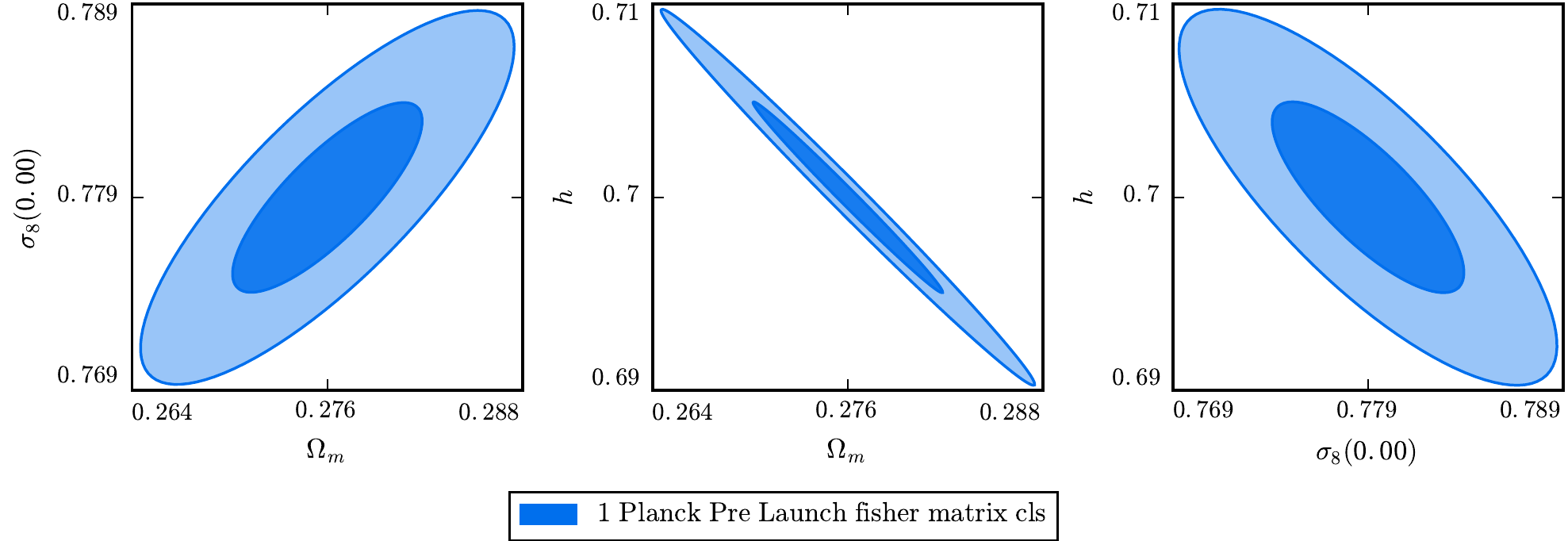} \label{SubFig:Planck2D} }
\hspace{5mm}
\subfigure[ {\it Planck} BB and DES+{\it Planck} BB forecasted constraints ]{ \includegraphics[width=0.9\textwidth]{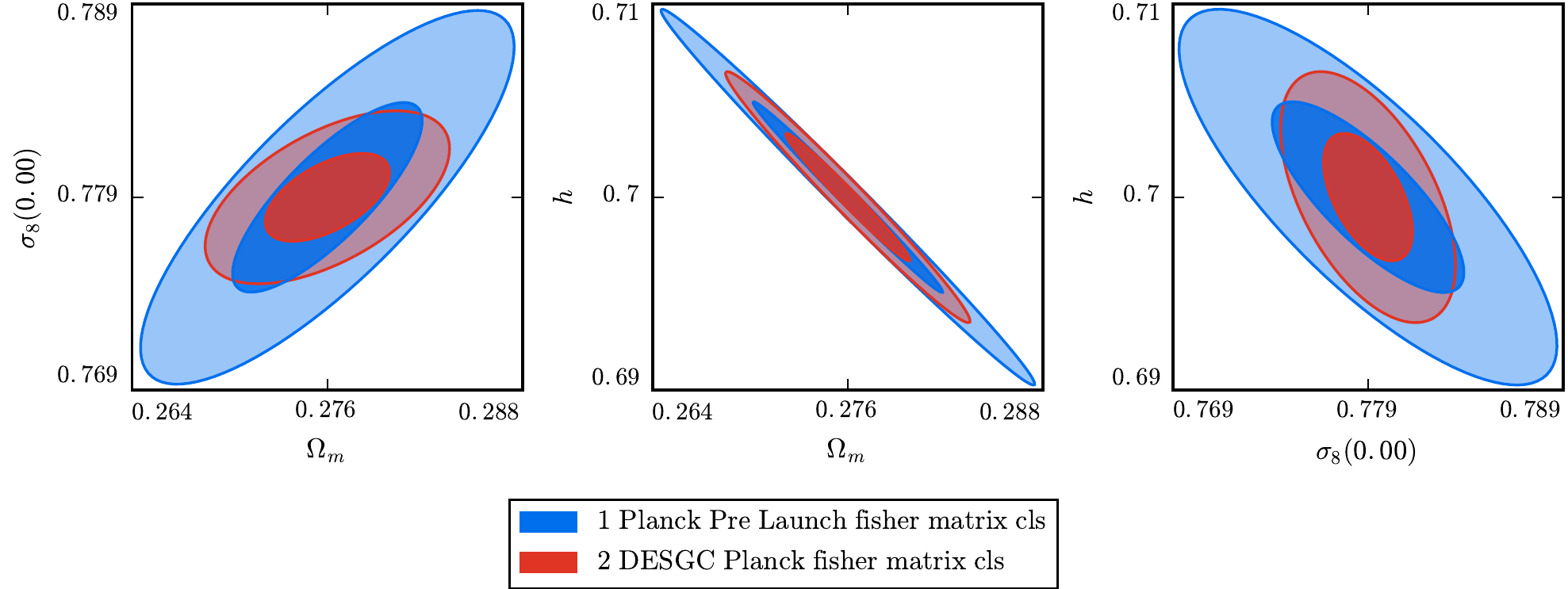}  \label{SubFig:DES2D} }
\caption{The marginalized 2D forecasted constraints on $\Omega_m$, $h$ and $\sigma_8$. Different colors correspond to different combinations of experiments, as shown in legend. The darker and lighter color shades indicate the forecasted $68\%$ C.L. and $95\%$ C.L. bounds.}\label{fig:2Dfigure}
\end{center}
\end{figure}

\begin{figure}[h!]
\begin{center}
\subfigure[ {\it Planck} BB forecasted constraints ]{ \includegraphics[width=13cm]{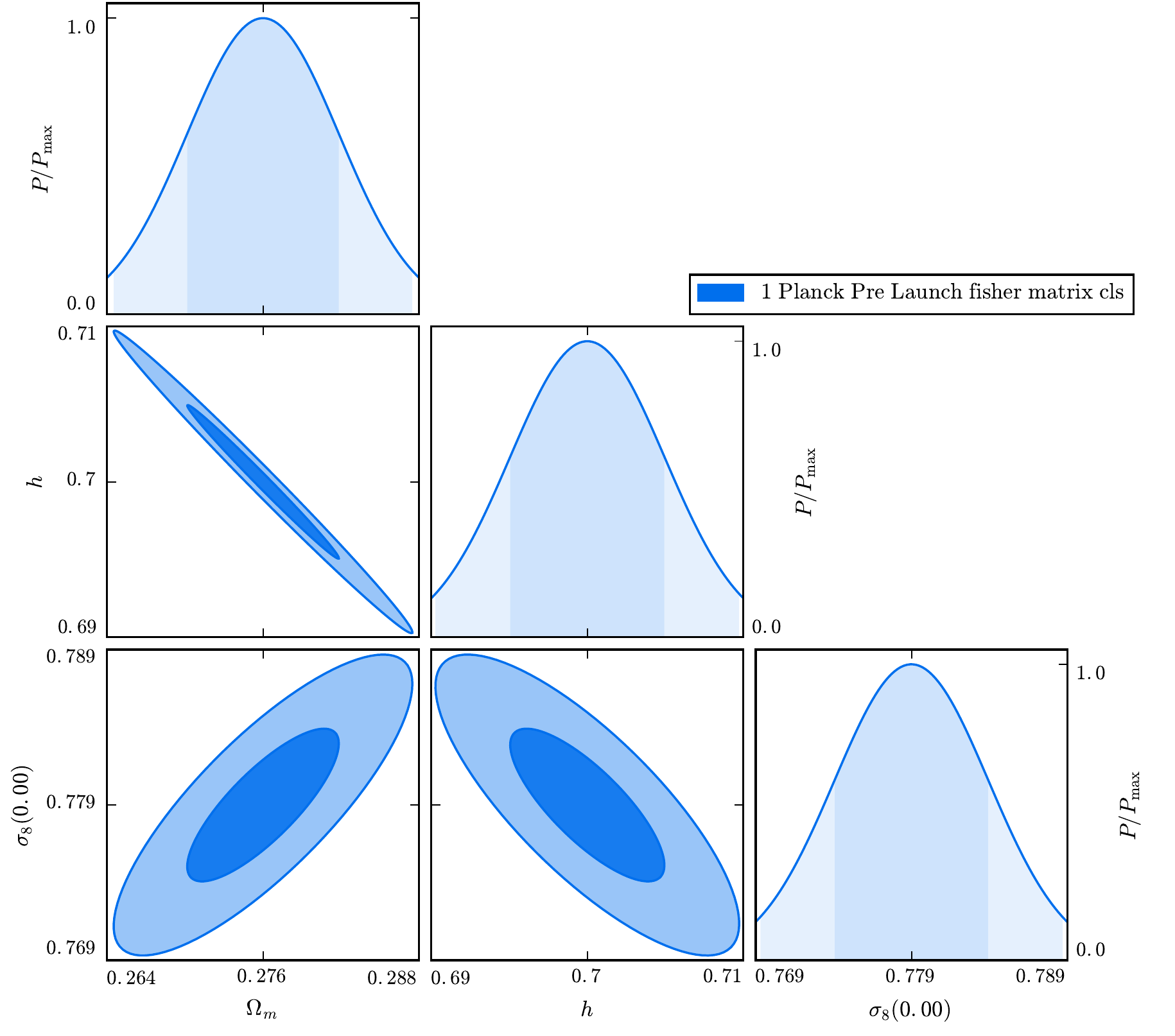} \label{SubFig:Plancktri} }
\hspace{5mm}
\subfigure[  {\it Planck} BB and DES+{\it Planck} BB forecasted constraints ]{ \includegraphics[width=13cm]{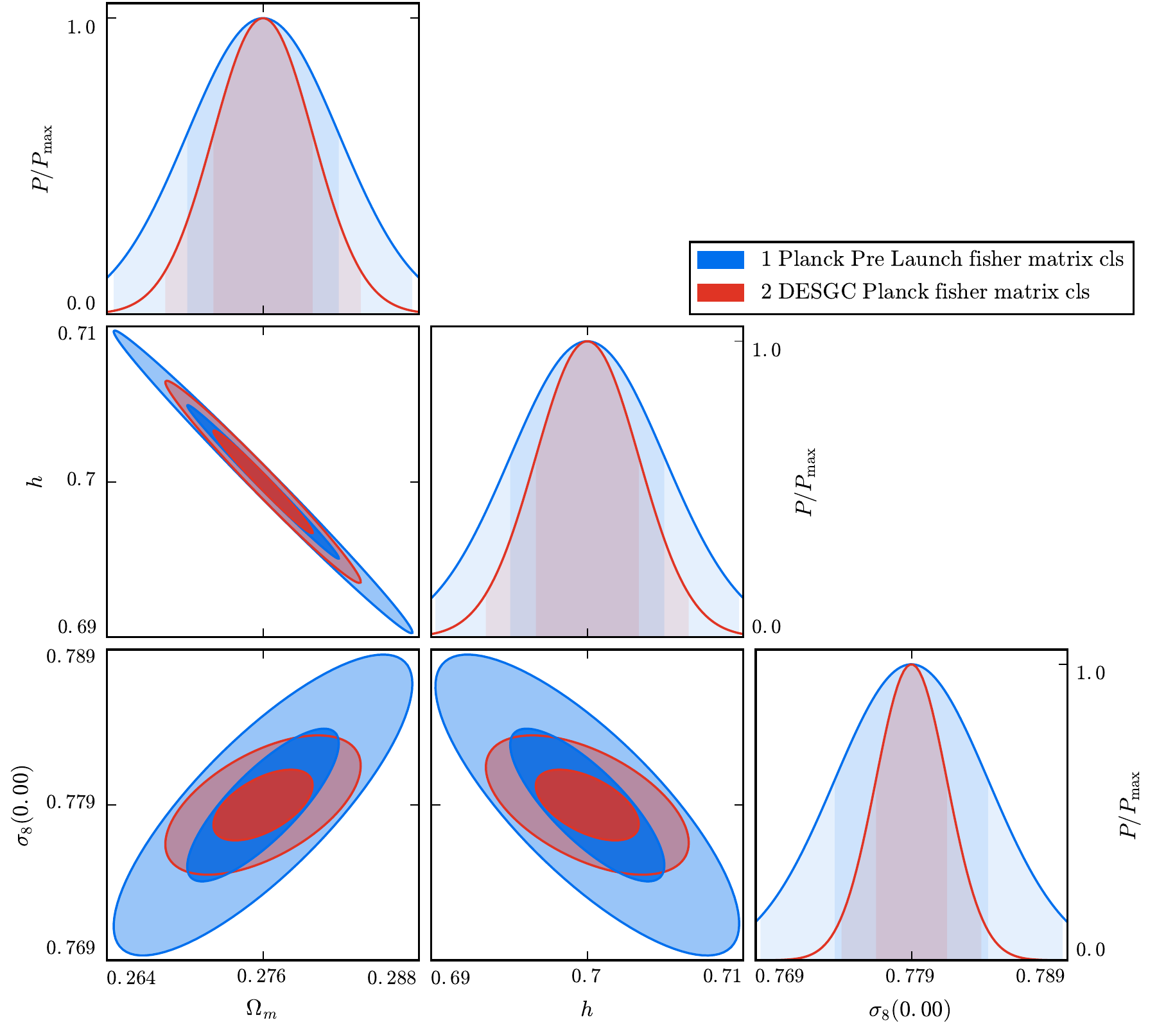}  \label{SubFig:DEStri} }
\caption{The triangular plot of the marginalized 2D and 1D forecasted constraints on $\Omega_m$, $h$ and $\sigma_8$. Different colors correspond to different combinations of experiments, as shown in legend. The darker and lighter color shades indicate the forecasted $68\%$ C.L. and $95\%$ C.L. bounds.}\label{fig:triplot}
\end{center}
\end{figure}

In this section we showcase some example usage of the CosmicFish code. All the results presented here are exactly the ones that can be obtained by running the example code distributed with the library.
To do so the user needs to go to the \code{examples} directory and issue the command \code{make all}, after compiling successfully the library.
This will produce all the Fisher matrices needed for these examples, by means of the Fortran CosmicFish library and then produce these plots with the Python CosmicFish library. \\
We choose to show, as examples, some of the forecast results for a {\it Planck} Blue Book (BB) CMB satellite, as taken from~\cite{bluebook}, and a DES-like Galaxy Clustering survey with specifications taken from~\cite{Lahav:2009zr}. \\
In Figure~\ref{fig:1Dfigure} we show the marginalized 1D plots on three cosmological parameters while in Figure~\ref{fig:2Dfigure} we show their 2D joint forecasted PDF. In Figure~\ref{fig:triplot} we show the triangular plot of these three parameters while Table~\ref{Tab:boundstable} shows the $68\%$ C.L. confidence forecasted bounds. In all figures we show the {\it Planck} BB constraints and the {\it Planck}+DES constraints to show an example of a plot displaying a the results from a single Fisher matrix and multiple ones. Notice that the code automatically maintains color consistency throughout all the plots, within a single execution of the program.

\begin{table}
\centering
\label{Tab:boundstable}
\begin{tabular}{ |l|l|l|l| }
\hline
\multicolumn{4}{|c|}{1 Planck Pre Launch fisher matrix cls marginal} \\[1mm]
\hline
 $\Omega_b h^2 = 0.0226 \pm 0.0001$ & $h = 0.7 \pm 0.005$                        & $n_s = 0.96 \pm 0.003$  & $\Omega_{m} = 0.276 \pm 0.006$       \\[1mm]
 $\Omega_c h^2 = 0.112 \pm 0.001$   & ${\rm{ln}}(10^{10} A_s) = 3.045 \pm 0.007$ & $\tau = 0.09 \pm 0.004$ & $\sigma_{8}(0.00) = 0.779 \pm 0.005$ \\[1mm]
\hline
\multicolumn{4}{|c|}{2 DESGC Planck fisher matrix cls marginal} \\[1mm]
\hline
 $\Omega_b h^2 = 0.0226 \pm 0.0001$ & $h = 0.7 \pm 0.003$                        & $n_s = 0.96 \pm 0.003$  & $\Omega_{m} = 0.276 \pm 0.004$       \\[1mm]
 $\Omega_c h^2 = 0.112 \pm 0.0007$  & ${\rm{ln}}(10^{10} A_s) = 3.045 \pm 0.005$ & $\tau = 0.09 \pm 0.003$ & $\sigma_{8}(0.00) = 0.779 \pm 0.002$ \\[1mm]
\hline
\end{tabular}
\caption{The $68\%$ confidence level bounds on cosmological parameters obtained with {\it Planck} BB and DES forecasts.}
\end{table}

 \newpage

\acknowledgments
We are grateful to Ana Ach\'ucarro, Carlo Baccigalupi, Erminia Calabrese, Stefano Camera, Luigi Danese, Giulio Fabbian, Noemi Frusciante, Bin Hu, Valeria Pettorino, Levon Pogosian, Giuseppe Puglisi and Alessandra Silvestri for useful and helpful discussions on the subject. We are indebted to Luca Heltai for help with numerical algorithms.
MM is supported by the Foundation for Fundamental Research on Matter (FOM) and the Netherlands Organization for Scientific Research / Ministry of Science and Education (NWO/OCW). MM was also supported by the DFG TransRegio TRR33 grant on The Dark Universe during the preparation of this work.
MR acknowledges partial support by the Italian Space Agency through the ASI contracts Euclid-IC (I/031/10/0) and the INFN-INDARK initiative.
MR acknowledges the joint SISSA/ICTP Master in High Performance Computing for support during the development of this work.
MR thanks the National Astronomical Observatories, Chinese Academy of Science for the hospitality during the initial phases of development of this work.
MR and MM thank the Galileo Galilei Institute for Theoretical Physics for the hospitality and the INFN for partial support during the completion of this work.
GBZ and YW are supported by the Strategic Priority Research Program ``The Emergence of Cosmological Structures'' of the Chinese Academy of Sciences Grant No. XDB09000000, and by University of Portsmouth. YW is supported by the NSFC grant No. 11403034. 


\end{document}